\documentclass[sigconf]{acmart}

\usepackage{enumitem}
\usepackage{algorithm}
\usepackage{algorithmic}
\usepackage{subfigure}
\usepackage{multirow}
\usepackage{CJKutf8}

\newcommand{\Real}{\mathbb{R}}

\newcommand{\Graph}{\mathcal{G}}

\AtBeginDocument{%
  \providecommand\BibTeX{{%
    \normalfont B\kern-0.5em{\scshape i\kern-0.25em b}\kern-0.8em\TeX}}}

\copyrightyear{2020}
\acmYear{2020}
\setcopyright{iw3c2w3}
\acmConference[WWW '20]{Proceedings of The Web Conference 2020}{April 20--24, 2020}{Taipei, Taiwan}
\acmBooktitle{Proceedings of The Web Conference 2020 (WWW '20), April 20--24, 2020, Taipei, Taiwan}
\acmPrice{}
\acmDOI{10.1145/3366423.3380077}
\acmISBN{978-1-4503-7023-3/20/04}

\begin{document}

\title{Beyond Clicks: Modeling Multi-Relational Item Graph for Session-Based Target Behavior Prediction}

\author{Wen Wang}
\email{51164500120@stu.ecnu.edu.cn}
\affiliation{%
  \institution{School of Computer Science and Technology,\\East China Normal University}
}

\author{Wei Zhang}\authornote{Wei Zhang is the corresponding author. This work is supported by NSFC (61702190), Shanghai Sailing Program (17YF1404500), and NSFC-Zhejiang (U1609220).}
\orcid{0000-0001-6763-8146}
\email{zhangwei.thu2011@gmail.com}
\affiliation{%
  \institution{School of Computer Science and Technology,\\East China Normal University}
}  

\author{Shukai Liu}
\author{Qi Liu}
\affiliation{\institution{Tencent}}
\email{shukailiu@tencent.com}
\email{addisliu@tencent.com}

\author{Bo Zhang}
\affiliation{\institution{Tencent}}
\email{nevinzhang@tencent.com}

\author{Leyu Lin}
\affiliation{\institution{Tencent}}
\email{	goshawklin@tencent.com}

\author{Hongyuan Zha}
\affiliation{\institution{Georgia Institute of Technology	}}
\email{zha@cc.gatech.edu}

\renewcommand{\shortauthors}{Wen Wang and Wei Zhang, et al.}

\begin{abstract}
Session-based target behavior prediction aims to predict the next item to be interacted with specific behavior types (e.g., clicking).
Although existing methods for session-based behavior prediction leverage powerful representation learning approaches to encode items' sequential relevance in a low-dimensional space, they suffer from several limitations.
Firstly, they focus on only utilizing the same type of user behavior for prediction, but ignore the potential of taking other behavior data as auxiliary information.
This is particularly crucial when the target behavior is sparse but important (e.g., buying or sharing an item).
Secondly, item-to-item relations are modeled separately and locally in one behavior sequence, and they lack a principled way to globally encode these relations more effectively. 
To overcome these limitations, we propose a novel Multi-relational Graph Neural Network model for Session-based target behavior Prediction, namely MGNN-SPred for short.
Specifically, we build a Multi-Relational Item Graph (MRIG) based on all behavior sequences from all sessions, involving target and auxiliary behavior types.
Based on MRIG, MGNN-SPred learns global item-to-item relations and further obtains user preferences w.r.t. current target and auxiliary behavior sequences, respectively.
In the end, MGNN-SPred leverages a gating mechanism to adaptively fuse user representations for predicting next item interacted with target behavior.
The extensive experiments on two real-world datasets demonstrate the superiority of MGNN-SPred by comparing with state-of-the-art session-based prediction methods, validating the benefits of leveraging auxiliary behavior and learning item-to-item relations over MRIG.
\end{abstract}




\maketitle

\section{Introduction}
Unlike conventional recommendation algorithms which get accustomed to modeling each user-item interaction separately~\cite{KorenBV09}, recent sequential recommendation approaches meet more realistic requirements for its ability of modeling user dynamic interest.
Session-based target behavior prediction~\cite{hidasi2015session} is the one of the main studied problem in this regard, aiming to predict the next item to be interacted with a user under a specific type of behavior (e.g., clicking an item).
Based on the predictions, information providers can effectively deliver items to appropriate users and at the same time, and users can quickly find the items what they actually want.
Note that we use session-based prediction and session-based recommendation interchangeably throughout this paper.

Early studies for this problem assume that the appearance of the next item depends only on its previous item~\cite{rendle2010factorizing,ZhangW15} in the same sequence.
With such a strong assumption, they could only model the last item in each sequence and ignore other information from the sequence.
To relieve this assumption, various methods adopt sequential models for session-based recommendation system to learn behavior sequences.
Recurrent Neural Networks (RNN)~\cite{Hochreiter-NC97} is commonly leveraged to obtain promising performance.
The relevant methods could roughly be attributed into two categories: single-session based recommendation models~\cite{hidasi2015session,gc_san} and multi-session based recommendation models~\cite{quadrana2017personalizing,YouWPERL19}. 
As the latter category requires the user ID of each behavior sequence should be known in advance to link multiple sequences of the same user together, it is not so universal than the first category due to privacy issues and user scalability problem (e.g., a billion of active users each day in WeChat). 
As such, we study session-based target behavior prediction from the perspective of single-session based modeling.

In the domain of single-session based behavior prediction, some studies~\cite{liu2018stamp,RenCLR0R19,sun2019bert4rec} adopt attention mechanism~\cite{Bahdanau-Arxiv14,Vaswani-NIPS2017} and outperform the pioneering RNN based methods~\cite{hidasi2015session}.
Recent advances in graph neural networks (GNN)~\cite{DefferrardBV16,hamilton2017inductive} further boost the performance of session-based behavior prediction by modeling each session-based behavior sequence as a graph to achieve the state-of-the-art performance~\cite{sr_gnn,gc_san}.
However, existing studies in this regard still suffer from several limitations.
\textit{Firstly}, they focus on only using the same type of user behavior as input for the next item prediction,
but ignore the potential of leveraging other type of behavior as auxiliary information.
This is particularly crucial when the target behavior is sparse but important (e.g., buying or sharing an item).
\textit{Secondly}, item-to-item relations are modeled separately and locally, since both RNN based and GNN based recommendation models only utilize one behavior sequence each time.
It is intuitive that abundant item-to-item relations are hidden in various behavior sequences.
For example, if many other users who have bought item B after buying item A, the relation between item A and B is especially vital if a target user just bought item A.

To overcome these limitations,
we propose a novel Multi-relational Graph Neural Network model for Session-based target behavior Prediction,
namely MGNN-SPred for short.
The target behavior we focused on is the aforementioned sparse behavior beyond the dense click behavior.
MGNN-SPred jointly considers target behavior and auxiliary behavior sequences and explores global item-to-item relations for accurate prediction.
Specifically, for the purpose of considering the global item-to-item relations,
we build a Multi-Relational Item Graph (MRIG) based on the past behavior sequences of all sessions.
There might exist multiple relations between two graph nodes, denoting target and auxiliary behavior types.
Based on MRIG, MGNN-SPred encodes global item-to-item relations into node representations and further obtains local representations for current target and auxiliary behavior sequences, respectively.
In the end, MGNN-SPred leverages a gating mechanism to adaptively fuse the representations from target behavior sequence and auxiliary behavior sequence to produce current user interest representation.

The main contributions of this work is summarized as follows:

1. We address the two limitations of existing methods by breaking the restriction of only using one type of behavior sequence in session-based recommendation and exploring another type of behavior as auxiliary information.
We further construct the multi-relational item graph for learning global item-to-item relations. 
    
2. To effectively model MRIG w.r.t. target and auxiliary behavior sequences, we develop the novel graph model MGNN-SPred which learns global item-to-item relations through graph neural network and integrates representations of target and auxiliary of current sequences by the gating mechanism.
    
3. We carry out extensive experiments and demonstrate MGNN-SPred achieves the best performance among strong competitors, showing the benefits of overcoming the two limitations.
As a byproduct, we release the source code\footnote{\url{https://github.com/Autumn945/MGNN-SPred}} of our model for relevant studies.

\section{Related Work}\label{sec:related}

\noindent\textbf{Session-Based Behavior Prediction.}
In the literature, the pioneering study~\cite{hidasi2015session} in the direction of single-session based recommendation first adopts a recurrent neural network based approach with past interacted items as the input of different time steps for session-based recommendation.
Following that, \cite{tan2016improved} improves the model with data augmentation and the consideration of temporal user behavior shift.
In addition to using RNN, \cite{li2017neural} also adopts attention mechanism to capture a user's sequential behavior and its main purpose in a current session.
Similarly, \cite{liu2018stamp} proposes a novel attention mechanism to capture both the users’ long-term interests in general and their short-term attention.
More recently, with the flourish Graph Neural Networks (GNN) methodologies, \cite{sr_gnn} first separates each session sequence into different graphs and uses graph neural networks to capture complex item transitions in a specific graph.
Afterwards, each session is represented as the combination of the global preference and current interests of this session using an attention network.
\cite{gc_san} is similar to \cite{sr_gnn}, which uses a multi-layered self-attention network as an alternative to capture long-range dependencies between items within a session.
As discussed in the introduction, these existing relevant methods suffer from two limitations which motivate the proposal of our model in this paper.

\vspace{.3em}\noindent\textbf{Multi-Behavior Modeling.}
Multi-behavior modeling for recommender system aims to leverage other types of user behavior to boost the recommendation performance on the target behavior.
A few studies have already investigated this scenario from different perspectives. 
~\cite{Krohn-GrimbergheDFS12} considers to leverage users' social interactions as auxiliary behavior for target behavior prediction by collective matrix factorization (CMF) techniques.
In a similar fashion,~\cite{ZhaoCHC15} builds multiple matrices from user different behaviors which cover user resharing behavior, user commenting behavior, user posting behavior, etc.
CMF is adopted to learn shared user representation for recommendation as well.
~\cite{LoniPLH16} proposes multi-feedback Bayesian personalized ranking (BPR), an extension of the classical Bayesian personalized ranking approach and tailored for different user behaviors.
It differentiates different preference levels between different user behaviors in the sampling stage for ranking.
~\cite{DingY0QLCJY18} also considers the assignment different preference levels of various user behaviors.
Instead of BPR, it incorporates this useful information into element-wise alternating least squares learner.
More recently, a neural network approach is proposed by~\cite{Gao0GCFLCJ19} to learn representations for user-item interactions with different behaviors. 
Multi-task learning is conducted to predict multi-behaviors with respect to a certain item in a cascading way.
Our work fundamentally differs from the above studies since all of them assume the independence of different user-item interactions while our study is more realistic by considering to model user behaviors in a sequential setting.

\vspace{.3em}\noindent\textbf{Graph Neural Networks.}
Graph neural networks are the methods used to generate representation of graph structured data, such as social network and knowledge graph.
\cite{perozzi2014deepwalk} extends Word2vec~\cite{mikolov2013distributed} by proposing a model, DeepWalk, to learn node representations based on sequences sampled from graphs.
LINE\cite{tang2015line} encodes first-order and second-order proximity of nodes into a low-dimensional space.
Recently, a surge of methods related on graph convolutional networks (GCN) have been raised.
\cite{bruna2013spectral} presents a method with a graph-based analogue of convolutional architectures, which is the original version of GCN.
Later, a number of improvements, extensions, and approximations of these spectral convolutions be proposed~\cite{kipf2016semi,duvenaud2015convolutional, hamilton2017inductive, monti2017geometric}.
These approaches outperform other methods based on random walks (e.g., DeepWalk and node2vec).
With the success in mind, an amount of GCN based methods are widely applied in various domains such as recommendation systems~\cite{monti2017geometric}.
But most GCN based methods require that all nodes in the graph are present in each propagation step of GNN.
Different from GCN, GraphSAGE~\cite{hamilton2017inductive} can train GNN with a minibatch setting.
Inspired by this, we design our GNN to learn from the constructed multi-relational item graph for session-based behavior prediction.

\begin{figure}
    \centering
    \includegraphics[width=0.5\textwidth]{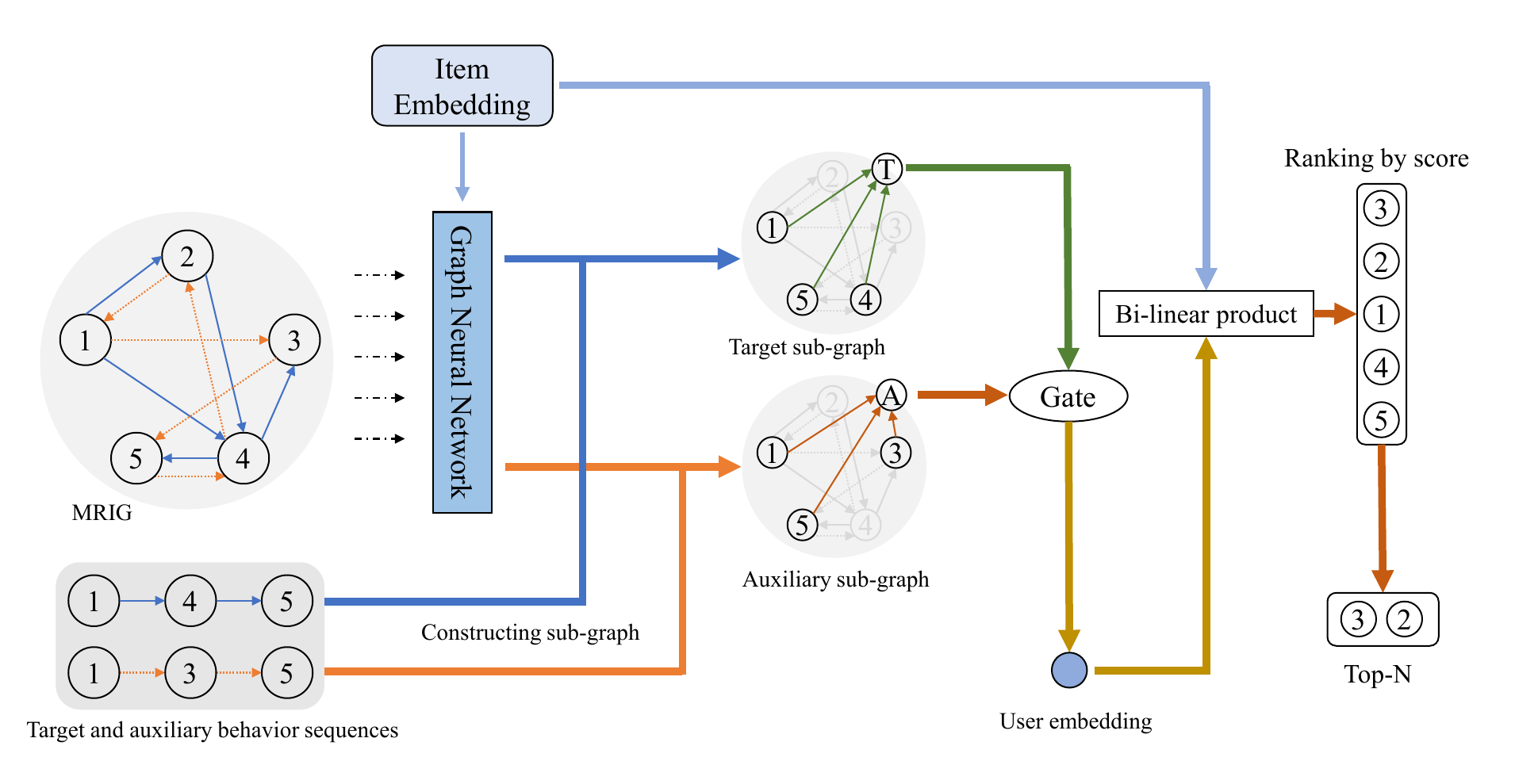}
	\vspace{-1.5em}
    \Description[Graph of model architecture]{Graph of model architecture containing input MRIG and output item ranking list.}
    \caption{The architecture of our model. We use a toy MRIG and two current behavior sequences as input. The number of recommended items is set to 2.}
    \label{fig:overview}
\end{figure}

\section{Technologies}\label{sec:method}

\subsection{Problem Definition}
For a session $s$ in the session set $S$, let $P^s = [p^s_1, p^s_2, p^s_3, ..., p^s_{|P^s|}]$ denote the target behavior sequence and $Q^s = [q^s_1, q^s_2, q^s_3, ..., q^s_{|Q^s|}]$ represent the auxiliary behavior sequence.
Moreover, we construct a Multi-Relational Item Graph $\Graph = (\mathcal{V}, \mathcal{E})$ based on all behavior sequences from all sessions,
where $\mathcal{V}$ is the set of nodes in the graph containing all available items and $\mathcal{E}$ is the edge sets involving multiple types of directed edges.
Each edge is a triple consisting of the head item, the tail item, and the type of this edge.
For instance, if we construct the graph based on behaviors of sharing and clicking, then an edge $(a, b, \textnormal{share}) \in \mathcal{E}$ means that a user shared item $a$ and subsequently shared item $b$,
and an edge $(a, b, \textnormal{click}) \in \mathcal{E}$ means that a user clicked item $b$ after clicking item $a$.
Given the above notations, we formulate the problem as follows:
\newtheorem{problem}{Problem}
\begin{problem}[session-based target behavior prediction]
Given a session $s \in S$ and its target and auxiliary behavior sequences $P^s$ and $Q^s$, along with MRIG $\Graph$,
the target of this problem is to learn a model that can generate $K$ items which are most likely to be interacted with the user of the session in the next.
\end{problem}

\begin{algorithm}[!t]
	\renewcommand{\algorithmicrequire}{\textbf{Input:}}
	\renewcommand{\algorithmicensure}{\textbf{Output:}}
	\caption{Multi-relational item graph construction}
	\label{alg:build_graph}
	\begin{algorithmic}[1]
		\REQUIRE
		Session set $S$,
		both target and auxiliary behavior sequences $P^s$ and $Q^s$, ${\forall}s \in S$
		\ENSURE
		MRIG $\Graph = (\mathcal{V}, \mathcal{E})$
		\STATE $\mathcal{V} \leftarrow \varnothing$, $\mathcal{E} \leftarrow \varnothing$
		\FOR{$s \in \mathcal{S}$}
            \STATE $\mathcal{V} \leftarrow \mathcal{V} \cup \{P^s[1]\}$
	        \FOR{$i = 2$ to $|P^s|$}
	            \STATE $\mathcal{V} \leftarrow \mathcal{V} \cup \{P^s[i]\}$, $\mathcal{E} \leftarrow \mathcal{E} \cup \{(P^s[i - 1], P^s[i], \textnormal{target})\}$
		    \ENDFOR
            \STATE $\mathcal{V} \leftarrow \mathcal{V} \cup \{Q^s[1]\}$
	        \FOR{$i = 2$ to $|Q^s|$}
	            \STATE $\mathcal{V} \leftarrow \mathcal{V} \cup \{Q^s[i]\}$, $\mathcal{E} \leftarrow \mathcal{E} \cup \{(Q^s[i - 1], Q^s[i], \textnormal{auxiliary})\}$
		    \ENDFOR
		\ENDFOR
	\end{algorithmic}

\end{algorithm}

\subsection{Overview}
The overall architecture of the proposed MGNN-SPred is depicted in Figure~\ref{fig:overview}.
The input to MGNN-SPred contains a Multi-Relational Item Graph (MRIG) and the two types of behavior sequences.
SR-MRIG first learns item correlations from MRIG by graph neural networks and encode them into item representations.
Afterwards, a user's two behavior sequences are regarded as two sub-graphs in the MRIG where the items in each sub-graph are connected with a virtual node (``T'' or ``A'' in Figure~\ref{fig:overview}), respectively.
Subsequently, SR-MRIG aggregates the nodes of each sub-graph to the corresponding virtual node, thus getting the representation of each behavior sequence.
Finally, to fuse the two behavior representations and obtain user preference representations,
a gating mechanism is adopted to adaptively decide the importance of different behaviors and perform weighted summation over them.
For the purpose of recommendation, SR-MRIG calculates each item's score by user and item representations via a bi-linear product and use the scores to rank them for recommendation.

\subsection{Graph Construction}
There are abundant relationships between items lying in users' historical behaviors.
If a user buys item $a$, and subsequently buys item $b$ in the same session,
it indicates that item $a$ and item $b$ probably have some dependency, but does not reflect similarity too much since a user less likely buys two very similar items within a short duration.
In comparison, if a user clicks item $a$, and subsequently clicks item $b$, it indicates that item $a$ and item $b$ are probably with large similarity.
This is intuitive because a user usually browses a number of similar items, and picks the most suitable one to buy.

We construct the multi-relational item graph by taking all items as nodes and each type of behavior corresponds as one directed edge, denoting different relationships between items.
The process of constructing MRIG is shown in Algorithm~\ref{alg:build_graph}.
The both target and auxiliary behavior sequences from all sessions $P^s$ and $Q^s$ (${\forall}s \in \mathcal{S}$) are provided as input.
The algorithm browses all behavior sequences, collects all items in the sequences as the nodes of the graph, and constructs edges between two consequent items in the same sequence with their behavior types as the edge types.
After constructing the graph with target and auxiliary behaviors, there are two types of directional edges in the graph.

\subsection{Item Representation Learning}
For each node $v \in \mathcal{V}$, we use $\bar{\mathbf{e}}_v \in \Real^{|\mathcal{V}|}$ denotes its one-hot representation.
Before we feed the one-hot representations of nodes into GNN, we first convert each of them into a low-dimensional dense vector $\mathbf{e}_v \in \Real^{d}$ by a learnable embedding matrix $\mathbf{E} \in \Real^{|\mathcal{V}| \times d}$: $\mathbf{e}_v = \mathbf{E}^\top\bar{\mathbf{e}}_v$.

After collecting the vectors $\mathbf{e}_v ({\forall}v \in \mathcal{V})$, we feed them with MRIG $\Graph$ into GNN to generate global representations of nodes $\mathbf{g}_v$.
The representations are expected to encode multiple item-to-item relations. 
We take node $v$ as an example for illustration. 
First of all, we collect neighbors of node $v$.
Each node in the graph has four types of neighboring node sets.
According to the type and direction, we name the four sets as ``target-forward'', ``target-backward'', ``auxiliary-forward'', and ``auxiliary-backward''. 
Take the type of ``target'' as an example,
we obtain neighbor groups corresponding to forward and backward directions as below:
\begin{equation}
    \mathcal{N}_{\textnormal{t}+}(v) = \{v' | (v', v, \textnormal{target}) \in \mathcal{E}\},~~
    \mathcal{N}_{\textnormal{t}-}(v) = \{v' | (v, v', \textnormal{target}) \in \mathcal{E}\}.
\end{equation}
For the type of ``auxiliary", its neighbor groups, i.e., $\mathcal{N}_{\textnormal{a}+}(v)$ and $\mathcal{N}_{\textnormal{a}-}(v)$, are acquired by the same way.

At each step of representation propagation in GNN, we first aggregate each group of neighbors by mean-pooling to obtain the representation of this group, defined as below:
\begin{equation}
	\mathbf{h}^k_{\textnormal{t}+, v} =  \frac{\sum_{v' \in \mathcal{N}_{\textnormal{t}+}(v)}\mathbf{h}^{k-1}_{v'}}{|\mathcal{N}_{\textnormal{t}+}(v)|}.
\end{equation}
The representations of the three remaining groups are calculated in a similar fashion. 
Consequently, for the propose of joint considering different relations between items,
we combine these four representations of different neighbor groups by sum-pooling:
\begin{equation}
	\mathbf{\bar{h}}^k_{v} =
	\mathbf{h}^k_{\textnormal{t}+, v} + 
	\mathbf{h}^k_{\textnormal{t}-, v} + 
	\mathbf{h}^k_{\textnormal{a}+, v} +
	\mathbf{h}^k_{\textnormal{a}-, v}.
\end{equation}
Finally, we update the representation of the center node $v$ by:
\begin{equation}
	\mathbf{h}^k_v = \mathbf{h}^{k-1}_v + \mathbf{\bar{h}}^k_v.
\end{equation}
After performing $K$ iterations, we take the node representation of the last step as the representation of the corresponding item: $\mathbf{g}_v = \mathbf{h}^K_v$.
In practice, we implement the GNN in a minibatch setting which is inspired by~\cite{hamilton2017inductive} to ensure scalability.

\subsection{Sequence Representation Learning}
We have tried different ways to compute the representation of the virtual node for the target and auxiliary behavior sequences, including using attention mechanism to assign different importance weights to the nodes and performing sub-graph propagating for several times.
Empirically, we have found that simple mean-pooling could already achieve comparable performance while retaining low complexity.
We denote the summarized representations of target behavior sequence $P$ and auxiliary behavior sequence $Q$ as $\mathbf{p}$ and $\mathbf{q}$, respectively, which are given as:
\begin{equation}
    \mathbf{p} = \frac{\sum_{i=1}^{|P|}\mathbf{g}_{p_i}}{|P|},~~\mathbf{q} = \frac{\sum_{i=1}^{|Q|}\mathbf{g}_{q_i}}{|Q|}.
\end{equation}

We argue that the two different types of behavior sequence representations might contribute differently when building an integrated representation.
This is because the auxiliary behavior is not exactly the same with the target behavior to be predicted, and different users might have different concentration on different behaviors.
For instances, some users might browse the item pages frequently and click various items arbitrarily, and another users might only click the items they want to buy.
It is self-evident that the contributions of auxiliary behavior sequence for the next item prediction are different in these situations. 
We define the following gating mechanism to calculate the relative importance weight $\alpha$:
\begin{equation}
    \alpha = \sigma(\mathbf{W}_g [\mathbf{p}; \mathbf{q}] ),
\end{equation}
where $[\mathbf{p}; \mathbf{q}]$ denotes the concatenation of the two representations, $\sigma$ is the sigmoid function, and $\mathbf{W}_g \in R^{1\times2d}$ is a trainable parameter of our model.
Finally, we obtain the user preference representation $\mathbf{o}$ for the current session by the weighted summation of $\mathbf{p}$ and $\mathbf{q}$:
\begin{equation}
    \mathbf{o} = \alpha \cdot \mathbf{p} + (1 - \alpha) \cdot \mathbf{q}.
\end{equation}

\subsection{Model Prediction and Training}
We further calculate the recommendation score $s_v$ of each item $v \in \mathcal{V}$ using the item embedding $\mathbf{e}_v$.
A bi-linear matching scheme is employed by:
\begin{equation}
    s_v = \mathbf{o}^\top \mathbf{W} \mathbf{e}_v,
\end{equation}
where $\mathbf{W} \in R^{d \times d}$ is a trainable parameter matrix of our model.

To learn the parameters of our model, we apply a softmax function to normalize the scores  $\mathbf{s} \in R^{|\mathcal{V}|}$ over all items to get the probability distribution $\hat{\mathbf{y}}$:
\begin{equation}
    \hat{\mathbf{y}} = \textnormal{softmax}(\mathbf{s}).
\end{equation}
Backpropagation for neural networks is adopted to optimize the model by minimizing the cross-entropy loss of the predicted probability distribution $\hat{\mathbf{y}}$ w.r.t. the ground truth.
The loss function is defined as follows:
\begin{equation}
    \mathcal{L}_{RS} = - \sum_{i = 1}^{|\mathcal{V}|} y_i \log(\hat{y}_i)\,,
\end{equation}
where $(y_1,\cdots,y_{|\mathcal{V}|})$ denotes the one-hot representation of the ground truth.
Note that $\mathcal{L}_{RS}$ is easily extended to a minibatch loss.

\section{Experiment}\label{sec:exp}

\begin{table}[!h]
\caption{Basic statistics of the datasets.}
\label{tab:dataset}
\vspace{-1em}
\resizebox{.8\columnwidth}{!}{
\begin{tabular}{lrr}
\hline
Data & WeChat & Yoochoose \\
\hline
\#items & 56,561 & 52,740 \\
\#sessions & 100,000 & 9,249,729 \\
Time duration & 2019/09/17\textasciitilde 23 & 2014/04/01\textasciitilde 09/30 \\
\#edge of target & 217,774  & 225,879 \\
\#edge of auxiliary & 1,546,220 & 3,277,411 \\
Average length of target & 9.76 & 3.31 \\
Average length of auxiliary & 33.49 & 8.56 \\
\#training data & 167,931  & 163,005 \\
\#validation data & 12,333   & 12,985 \\
\#test data & 24,667   & 25,971 \\
\hline
\end{tabular}
}
\end{table}

\subsection{Experimental Setup}

\subsubsection{Dataset}
We evaluate our model on two real-world datasets named WeChat and Yoochoose.
The Yoochoose dataset is obtained from the RecSys Challenge 2015.
The user behavior sequences in the dataset are already segmented into sessions and all the users are anonymized.
The WeChat dataset is collected from \emph{Top Stories} (\begin{CJK}{UTF8}{gbsn}看一看\end{CJK}) of WeChat, where we choose videos are regarded as items.
We randomly select one hundred thousand active users and collect their behavior records for a duration of one week.
Since the duration is relatively short, we retain an entire behavior sequence of each user by taking the sequence as a single session.
In this paper, we treat the behavior of purchase in Yoochoose and the behavior of sharing in WeChat as the target behavior and regard the behavior of clicking in both datasets as the auxiliary behavior.

Given a session with the target behavior sequence $P = [p_1, p_2, ..., \\p_{|P|}]$ and the auxiliary behavior sequence $Q = [q_1, q_2, ..., q_{|Q|}]$, we adopt a similar way to construct training example as~\cite{li2017neural, sr_gnn}.
That is, we treat each item $p_i$, ($i \ge 2$) as the label and use $[p_1, p_2, ...p_{i-1}]$ as input of target behavior.
The treatment for the auxiliary behavior is a little different, because a user is very likely to click an item before buying or sharing it.
To avoid the auxiliary input already sees the labels, we only keep the clicked items before the target item that is also bought or shared by the user.
We set a maximum length $L$ for both types of sequences and only keep the last $L$ items longer than the maximum length.
Considering the fact that two datasets have different average sequence length (see details in Table~\ref{tab:dataset}), we set the maximum length $L$ to 10 for WeChat and 3 for Yoochoose.
We discuss the impact of different maximum length in Section~\ref{sec:len}.

We split the datasets in a chronological order for evaluation, consistent with real situations.
We take the first 6/7 of datasets as the training data, and use 1/3 of the remaining data as the validation data to determine optimal hyper-parameter settings.
MRIG used throughout the experiments are constructed only based on training data.
The basic statistics of two datasets are summarized in Table~\ref{tab:dataset}.

\subsubsection{Baselines}
We compare the proposed model with several strong competitors, including state-of-the-art graph neural network based model for session-based recommendation.

\begin{itemize}[leftmargin=*]
	\item \textbf{POP.} It just recommends the top-n frequent items in the training set regardless of behaviors in current sessions. 
	\item \textbf{Item-KNN~\cite{sarwar2001item}.} It recommends items most similar to the previously interacted items belonging to the same sessions.
	\item \textbf{GRU4Rec~\cite{hidasi2015session}.} GRU4Rec is the pioneering RNN-based deep sequential model for session-based recommendation.
	\item \textbf{NARM~\cite{li2017neural}.} It employs attention mechanism to capture different importance of each item according to their hidden states obtained by RNN.
	A weighted integration of different item representations is performed to obtain final representation.
	\item \textbf{STAMP~\cite{liu2018stamp}.} This model learns users’ general interest from the long-term memory of session context and current interest from the short-term memory of their last behaviors.
	\item \textbf{SR-GNN~\cite{sr_gnn}} and \textbf{GC-SAN~\cite{gc_san}.}
    Both of the graph-based models only use a current session to construct graph for applying GNN to learn item representations.
    The difference is SR-GNN represents each session by a traditional attention network while GC-SAN is based on a multi-layered self-attention mechanism.
	
	\item \textbf{R-DAN.}
    Reasoning-DAN (R-DAN)~\cite{nam2017dual} is used to model both behavior sequences simultaneously.

	\item \textbf{CoAtt.}
    Co-Attention (CoAtt)~\cite{lu2016hierarchical} with alternative calculation for interactive attention is adopted for comparison.
    
	\item \textbf{HetGNN.}
	Heterogeneous graph neural network~\cite{ZhangSHSC19} is applied for recommendation, with two edge types and one node type.

\end{itemize}

It is worth noting only target behavior is considered by the above baselines originally developed for session-based recommendation, i.e., GRU4Rec, NARM, STAMP, SR-GNN, and GC-SAN.
To make the comparison more fairable, we revise these methods through the following manner.
We use their original forms to model the target behavior sequence and auxiliary behavior sequence respectively,
And afterwards, we utilize the proposed gating mechanism to fuse the two types of representations as ours. 
In addition, we also compare our model with the baselines in the situation of only considering target behavior (see Table~\ref{tab:click_result} for details).

\begin{table}[!t]
\caption{Evaluation results of all methods.}
\centering
\label{tab:result}
\vspace{-1em}
\resizebox{\columnwidth}{!}{
\begin{tabular}{lccccccc}
\hline
\multirow{2}{*}{Methods} & \multicolumn{3}{c}{WeChat} & & \multicolumn{3}{c}{Yoochoose} \\
\cline{2-4}
\cline{6-8}
 & H@100 & M@100 & N@100 & & H@100 & M@100 & N@100 \\
\hline
\hline
POP     & 13.565 & 1.1247 & 3.2621 & & 6.095  & 0.2529 & 1.2231 \\
Item-KNN& 15.770 & 1.1624 & 3.7222 & & 15.286 & 1.9415 & 4.4040 \\
\hline
GRU4Rec & 18.831 & 1.3956 & 4.3966 & & 19.114 & 2.5292 & 5.5830 \\
NARM    & 19.131 & 1.4034 & 4.4416 & & 18.775 & 2.5819 & 5.5813 \\
STAMP   & 17.757 & 1.3083 & 4.1078 & & 20.361 & 2.3487 & 5.6879 \\

\hline
SR-GNN  & 18.940 & 1.3827 & 4.3967 & & 21.262 & 2.6892 & 6.1232 \\
GC-SAN  & 19.034 & 1.2090 & 4.2490 & & 19.718 & 2.5218 & 5.6861 \\
HetGNN & 20.290 &  1.4171 & 4.6504 && 24.031 & 2.9546 & 6.8732 \\
\hline
R-DAN     & 18.952 & 1.3879 & 4.3980 & & 15.956 & 2.3107 & 4.8608 \\
CoAtt   & 17.700 & 1.1931 & 4.0137 & & 20.080 & 2.5742 & 5.8206 \\
\hline
\textbf{Ours} & \textbf{21.271} & \textbf{1.4797} & \textbf{4.8529} & & \textbf{28.632} & \textbf{3.6564} & \textbf{8.2722} \\
\hline
\end{tabular}
}
\end{table}

\begin{table}
\caption{Results of not using auxiliary behavior sequences.}
\centering
\label{tab:click_result}
\vspace{-1.em}
\resizebox{\columnwidth}{!}{
\begin{tabular}{lccccccc}
\hline
\multirow{2}{*}{Methods} & \multicolumn{3}{c}{WeChat} & & \multicolumn{3}{c}{Yoochoose} \\
\cline{2-4}
\cline{6-8}
 & H@100 & M@100 & N@100 & & H@100 & M@100 & N@100 \\
\hline
\hline
GRU4Rec (w/o a) & 16.889 & 1.2346 & 3.9128 & & 14.817 & 1.6032 & 4.0012 \\
NARM (w/o a)    & 17.773 & 1.3123 & 4.1298 & & 14.443 & 1.5540 & 3.8900 \\
SR-GNN (w/o a)  & 18.093 & 1.2621 & 4.1368 & & 15.302 & 1.5782 & 4.0852 \\
Ours (w/o a)    & 19.252 & 1.3933 & 4.4473 & & 21.089 & 2.3798 & 5.8221 \\
\hline
\end{tabular}
}
\end{table}

\subsubsection{Implementation Details}
We implement our proposed model based on Tensorflow.
The dimension of item embedding is set to 64.
Adam with default parameter setting is adopted to optimize the model, with the mini-batch size of 64.
GNN is ensured to run in a minibatch setting and the depth $K$ is set to 2.
We terminate the learning process with an early stopping strategy.
We test different forms of attention computation formulas for the baselines based on attention mechanism and report their best results.
The hyper-parameters of baselines are turned on validation datasets as well.

\subsection{Model Comparison}
We consider the top-100 ranked predictions as recommended items.
Following~\cite{sr_gnn, gc_san}, we adopt HR@100 (H@100), MRR@100 (M@100), and NDCG@100 (N@100) to evaluate the recommendation performance of all models after obtaining their recommendation lists.
Table~\ref{tab:result} shows the performance comparison between our model and the adopted baselines.
(1) The first part of the table corresponds to the simple baselines.  
We observe their results are significantly worse than other methods.
(2) The second part involves standard sequential based methods for session-based recommendation.
We observe that their results keep at the same level, except for STAMP on WeChat.
It shows that: 1) taking session-based recommendation as a sequential modeling task can improve performance; 2) although NARM and STAMP are more advanced approaches which use attention mechanism to combine hidden representations of different time steps, they do not show advantages on the sparse behavior prediction problem we studied (not the same as previous studies focusing on click prediction).
(3) The third part is GNN based models. 
SR-GNN and GC-SAN seem to be better than the sequential methods, and HetGNN further boost the performance.
(4) The second-to-last part involves approaches of learning two sequences in other research domains.
Their best results are worse than the best performance of the above recommendation methods,
which suggests that considering the interaction of items in two sequences might have no benefit for the studied problem.
Finally, we can see that our method outperforms all the other methods, demonstrating the superiority of our model for session-based recommendation.

\subsection{Impact of Auxiliary Behavior Sequence}
We choose several representative methods in Table~\ref{tab:click_result} to test whether considering the auxiliary behavior sequence indeed boosts the performance of session-based recommendation.
The methods with ``(w/o a)" mean removing the auxiliary behavior sequence from their full version.
Firstly, we observe that our proposed model still consistently achieves better performance in this situation.
Moreover, by comparing each method in Table~\ref{tab:result}  with its ``(w/o a)" version, we can find every method beats the one of ``(w/o a)" with significant margins. 
Based on the above illustrations, we demonstrate that considering the auxiliary behavior sequence is indeed meaningful.

\begin{table}[!t]
\centering
\caption{Ablation study of MGNN-SPredl.}
\label{tab:ablation}
\vspace{-1.em}
\resizebox{\columnwidth}{!}{
\begin{tabular}{lccccccc}
\hline
\multirow{2}{*}{Methods} & \multicolumn{3}{c}{WeChat} & & \multicolumn{3}{c}{Yoochoose} \\
\cline{2-4}
\cline{6-8}
 & H@100 & M@100 & N@100 & & H@100 & M@100 & N@100 \\
\hline
\hline
Ours (w/o ae)      & 20.923 & 1.4665 & 4.7945 & & 25.463 & 2.7678 & 6.8907 \\
Ours (w/o asg) & 19.742 & 1.3949 & 4.5167 & & 22.517 & 2.6025 & 6.2631 \\
Ours (w/o g)            & 20.363 & 1.3707 & 4.6154 & & 27.577 & 3.3531 & 7.7896 \\
\hline
\textbf{Ours} & \textbf{21.271} & \textbf{1.4797} & \textbf{4.8529} & & \textbf{28.632} & \textbf{3.6564} & \textbf{8.2722} \\
\hline
\end{tabular}
}
\end{table}

\subsection{Model Analysis}
\subsubsection{Ablation Study}
We conduct ablation studies of our model, using ``w/o ae" to denote removing the edges related to the auxiliary behavior, using ``w/o asg" to denote that not modeling the sub-graph of the auxiliary behavior sequence in getting user preference representation, and using ``w/o g" to indicate merging the two representations of the target and auxiliary behavior sequences by simple summation instead of the gating mechanism.
Table~\ref{tab:ablation} shows the corresponding results.
We observe that the incorporation of the auxiliary edge into the built graph is beneficial for the problem by seeing ``w/o ae".
The integration of the auxiliary behavior with target behavior sequence have a notable contribution by seeing ``w/o asg".
Besides, we find that the performance becomes worse if we do not use the gating mechanism to merge the two representations of the target and auxiliary behavior sequences by investigating ``w/o g".
Through the above comparison, we conclude the main components in our model are effective.

\begin{figure}[!h]
	\center
	\subfigure[\small{WeChat}]
	{\label{fig:wx_depth}
		\includegraphics[width=0.225\textwidth]{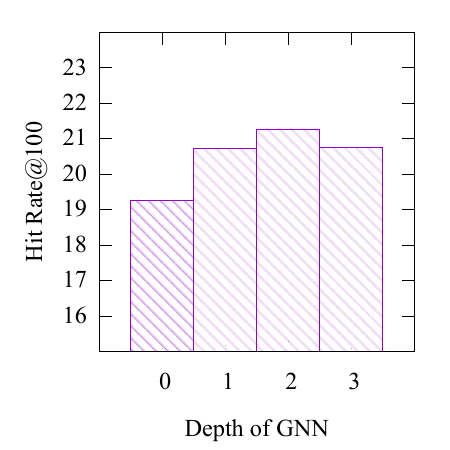}
	}
	\subfigure[\small{Yoochoose}]
	{\label{fig:yc_depth}
		\includegraphics[width=0.225\textwidth]{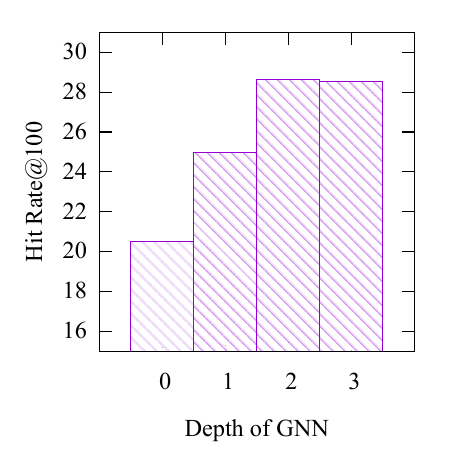}
	}
	\vspace{-1.em}
	\Description[Histogram of results]{Histogram of results corresponding to different versions of our model with with different number of layers in GNN.}
	\caption{Results of our model with different depths of GNN.}
	\label{fig:depth}
\end{figure}

\subsubsection{Impact of Depth of GNN}
We test different depth settings (from 0 to 3) about graph representation propagation.
The depth setting with value 0 means the our model does not use GNN and could not learn any information from MRIG.
Figure~\ref{fig:depth} shows the corresponding results.
We can see that the performance of depth 0 is without doubt much worse than the results with depths from 1 to 3.
This comparison clarifies the significance of considering MRIG for our model.
Moreover, the performance becomes significantly better when the depth grows from 1 to 2, showing modeling high-order relation between items through GNN is indispensable. 
When the number of graph representation propagation is larger than 3, the representations of nodes might become less distinguishable, which is not ideal for further improving the performance.

\begin{figure}[!t]
    \vspace{-1.em}
	\center
	\subfigure[\small{WeChat}]
	{\label{fig:wx_len}
		\includegraphics[width=0.225\textwidth]{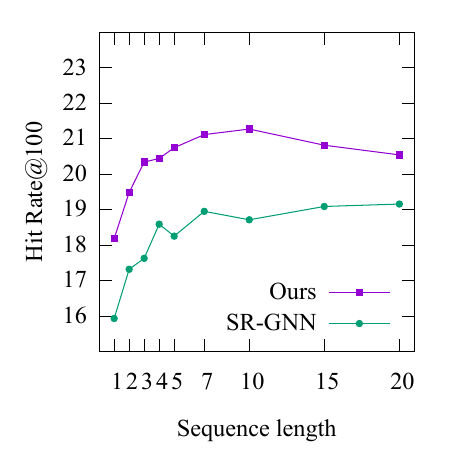}
	}
	\subfigure[\small{Yoochoose}]
	{\label{fig:yc_len}
		\includegraphics[width=0.225\textwidth]{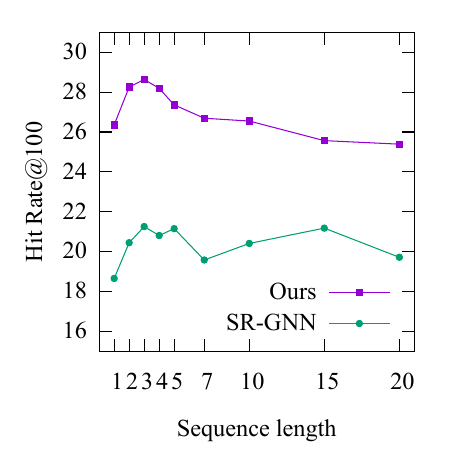}
	}
	\vspace{-1.em}
	\Description[Line charts]{Line charts of different results corresponding to different sequence lengths.}
	\caption{Results for different maximum lengths.}
	\label{fig:length}
\end{figure}

\subsubsection{Impact of Sequence Length}
\label{sec:len}
We visualize the performance variation with the change of the maximum behavior sequence length $L$ in Figure~\ref{fig:length}, where we set $L$ in the range from 1 to 20.
As expected, with larger maximum sequence length at the beginning, the performance of both our model and SR-GNN grows to be better.
After reaching the peaks, the results slightly become worse, and finally the variation trends turn to be stable.
Overall, our model outperforms SR-GNN consistently.
Besides, we find the lengths with the best performance are not the same in the two datasets. 
This is due to the fact the average length of Yoochoose is much smaller than that of WeChat, as shown in Table~\ref{tab:dataset}.

\section{Conclusion}\label{sec:con}
In this paper, we study session-based target behavior prediction.
Two limitations of existing relevant models are addressed: using only target behavior for next item prediction and lacking a principled way to encode global item-to-item relations.
To alleviate the issues, MGNN-SPred is proposed, with the major novelties of building and modeling of the multi-relational item graph.
In addition, a gating mechanism is adopted to adaptively fuse target behavior sequences and auxiliary behavior sequences into the user preference representations for the next item prediction. 
Comprehensive experiments on two real-world datasets demonstrate MGNN-SPred achieves the best performance and its design is rational.


\bibliographystyle{ACM-Reference-Format}
\bibliography{acmart}

\end{document}